\documentclass[aps,prl,twocolumn,showpacs,floatfix,superscriptaddress,tightenlines,amsmath,amssymb,yhmath]{revtex4-2}
\usepackage{hyperref}
\usepackage[dvipsnames,x11names]{xcolor}
\pagecolor{white}
\usepackage[final]{graphicx}
\usepackage{bm}
\usepackage{float}
\usepackage{multirow}
\usepackage{enumerate, enumitem}
\usepackage[normalem]{ulem}
\usepackage[capitalize]{cleveref}
\usepackage[utf8]{inputenc}

\newcommand{\Vvec}{{\bm V}}
\newcommand{\bvec}{{\bm b}}
\newcommand{\A}{\mathcal{A}}
\newcommand{\B}{\mathcal{B}}

\newcommand{\evec}{\bm e}

\newcommand{\uvec}{{\bm u}}

\newcommand{\Gz}{{\Gamma_0}}
\newcommand{\Gt}{{\Gamma_2}}
\newcommand{\vo}{v_0}
\newcommand{\xvec}{{\bm x}}
\newcommand{\Rey}{\textrm{Re}}

\crefformat{equation}{(#2#1#3)}
\crefrangeformat{equation}{(#3#1#4) to~(#5#2#6)}
\crefmultiformat{equation}{(#2#1#3)}%
{ and~(#2#1#3)}{, (#2#1#3)}{ and~(#2#1#3)}

\begin{document}

\title{Flocking and mesoscale turbulence in three-dimensional active fluids}
\author{Prasad Perlekar}
\email{perlekar@tifrh.res.in}
\affiliation{Tata Institute of Fundamental Research, 36/P, Gopanpally Village, Serilingampally Mandal, Ranga Reddy District, Hyderabad 500046, Telangana, India}%

\begin{abstract}
We numerically study the three-dimensional turbulence in a minimal model of an active fluid--the Toner-Tu-Swift-Hohenburg equation. For small activity, we observe bacterial turbulence, while for large activity, we uncover hitherto unexplored regime of a turbulent flock where a global order coexists with turbulence. We present a simple closure model that predicts the turbulent flock and also qualitatively explains the transition to the bacterial turbulence regime via a transcritical bifurcation. 
\end{abstract}

\maketitle
 A collection of organisms, active matter, shows spectacular patterns and  dynamics on scales much larger than the size of an individual  \cite{ramaswamy2010,marchetti2013, chate2020,ramaswamy2017,ramaswamy2019,cates2019,bees2020advances}. The two commonly encountered but contrasting phases of active matter are: (a) Flocking: global-ordered motion emerging from local alignment interactions between agents, as observed in bird flocks \cite{vicsek1995,toner2005,chate2008a,cavagna2014, maitra2020,chatterjee2021,rana2022,jain2024,jain2025}; and (b) Turbulence: complex spatiotemporal flows featuring coherent vortices and jets generated by steric interactions and disturbances from individual agents, as seen in bacterial suspensions  and microtubule-kinesin mixtures \cite{dunkel2013,wensink2012,giomi2015,doostmohammadi2018,rana2022,james2021, mukherjee2023,jain2024,jain2025,kiran2025}.
 
 The hydrodynamic theory of active matter provides an effective description of these phases \cite{ramaswamy2010,chate2020}. In the Stokesian limit, a key prediction is that long-range flocking cannot be sustained in an unbounded fluid \cite{simha2002}. Indeed, dense suspensions of motile micro-swimmers stir the fluid with their flagella and exhibit collective behavior described as mesoscale turbulence \cite{dombrowski2004,wensink2012,dunkel2013,guo2018,liu2021,gautam2024,wei2024}. The Toner-Tu-Swift-Hohenberg equation (TTSH) \cite{wensink2012,dunkel2013,bratanov2015,sanjay2020,puggioni2023,mukherjee2023} and the Generalized Navier-Stokes equation (GNS) \cite{slomka2015,slomka2017,linkmann2019,linkmann2020} serve as minimal continuum models that quantitatively capture the dynamics of the bacterial and solvent velocity fields, respectively. Beyond enabling comparison with experiments, these equations offer a natural framework for studying pattern formation and turbulence in active matter. Numerical studies indicate that turbulence in both the GNS and TTSH models closely resembles canonical fluid turbulence and, over a range of parameters, exhibits an inverse energy cascade in two dimensions \cite{mukherjee2023,puggioni2023,linkmann2019}. More intriguingly, the 3D GNS equations allow for Beltrami flows and the emergence of an inverse energy cascade \cite{slomka2015,slomka2017}, which cannot be observed in the Navier-Stokes (NS) equations unless the chiral symmetry is broken by using a numerical methodology \cite{biferale2012}.  
 
 An early experimental study \cite{dunkel2013}, characterized the statistical properties of three-dimensional (3D) bacterial flows and described them using the TTSH framework, a comprehensive study of collective motion in 3D bacterial suspensions  is still lacking. Recent experiments have systematically examined the transition from two- to three-dimensional turbulence by varying the degree of confinement \cite{guo2018,wei2024}. These investigations demonstrate that, beyond a critical confinement, the characteristic size of coherent bacterial flow structures grows with the square root of the confinement length. These observations behoove us to investigate whether a global order can emerge in an unbounded bulk bacterial suspension. 

We address this question within the framework of the minimal model for the bacterial velocity field, namely the incompressible Toner–Tu–Swift–Hohenberg (TTSH) equation:
\begin{align}
\begin{split}
\partial_t \uvec + \lambda \uvec \cdot \nabla \uvec =& -\nabla P - \Gz \nabla^2 \uvec - \Gt \nabla^4 \uvec \\
&+ (\alpha - \beta |\uvec|^2) \uvec, 
\end{split}
\label{eq:ttsh}
\end{align}
where $\uvec$ is the coarse-grained velocity field, $\Gz$ controls the strength of the small-scale stirring, $\Gt$ models small-scale dissipation, $P$ is the pressure that enforces the incompressibility criterion $\nabla \cdot \uvec=0$, and we choose the coefficient of the cubic damping term $\beta>0$. The time scale for the injection of energy due to stirring is $\tau = \Gt/\Gz^2$. In steady state, a homogeneous suspension is at rest $\uvec={\bm 0}$  for $\alpha<0$ or moves at a uniform speed $\uvec={\bm U}$ with $U\equiv \sqrt{\alpha/\beta}$ for $\alpha>0$. Refs.~\cite{heiden2016,reinken2018} obtain hydrodynamic equations for swimmer suspensions from a generic microscopic model and demonstrate that the TTSH equations arise in the limit where the solvent velocity does not depend on the swimmer orientation.

\begin{figure*}[!ht]
\includegraphics[width=0.49\linewidth]{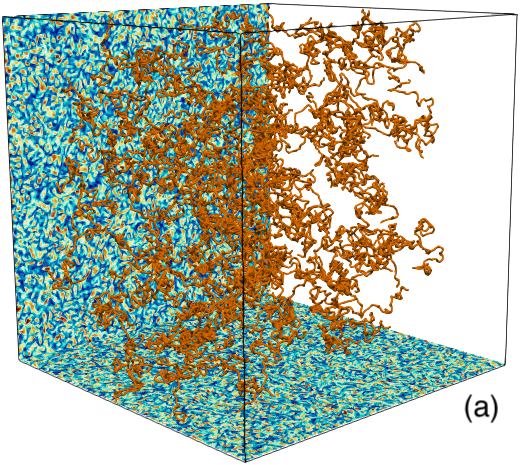}
\includegraphics[width=0.47\linewidth]{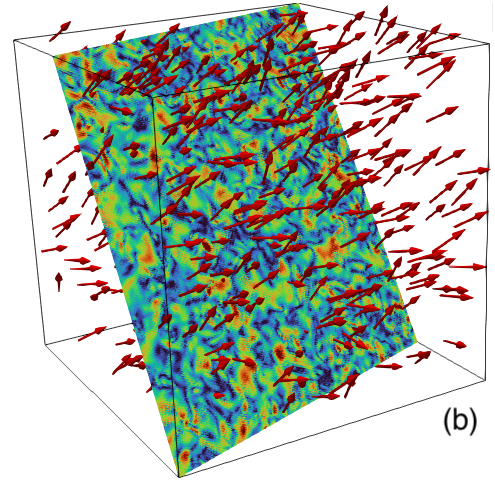}
\caption{\label{fig:f1} (a) Mesoscale turbulence [$\A=0.44,\lambda=1.7$ (run {\tt R1})]: Streamlines of the velocity field, and the pseudo-color plot of the magnitude of the vorticity field in the XY and XZ plane. (b) Turbulent flock [$\A=1.1,\lambda=1.7$ (run {\tt R1})]: Velocity vectors showing large scale ordering, and two dimensional slice of the magnitude of the vorticity field in a plane perpendicular to the average velocity shows turbulence-like flow structures.}
\end{figure*}

We perform direct numerical simulations (DNS) of the TTSH equations using a pseudo-spectral method in a three-dimensional periodic cubic domain with each side of length up to $L=96 \pi$ discretized with $N^3=1024^3$ collocation points. Integration of \cref{eq:ttsh} is  performed by means of a fully dealiased pseudospectral code, with the second-order Adams-Bashforth scheme with exactly integrated viscous term.  All simulations are performed for at least $t\approx 10^3\tau$ and a statistically steady state is obtained after $t \approx 20 \tau$. The coefficients $\beta=0.1$, $\Gamma_0=0.9$, and $\Gamma_2=0.9$ are kept fixed and chosen to be consistent with previous experiments and simulations designed to study mesoscale turbulence in 3D  \cite{dunkel2013}. \cref{table} summarizes the parameters used in our simulation. The flow is characterized in terms of three dimensionless parameters: the coefficient of advective nonlinearity $\lambda$, the activity parameter ${\mathcal A}\equiv \alpha \tau$, and $\B=\beta \Gz$ (see End Matter). In what follows, we vary $\lambda$ and $\A$ and investigate the flow properties in a statistically steady state.

Our study uncovers a novel nonequilibrium phase transition from mesocale turbulence to an ordered but fluctuating flock (large scale order in turbulent background). From \cref{eq:ttsh}, we construct a reduced-order model for mean velocity and fluctuating kinetic energy, which quantitatively reproduces the fluctuating flock state and also qualitatively captures an activity-controlled transcritical bifurcation between the two states.

In \cref{fig:f1}(a), we show the streamlines of the velocity field together with pseudocolor visualizations of the magnitude of the vorticity field in the XY and XZ planes for $\A=0.44$ and $\lambda=1.7$. Consistent with previous studies \cite{dunkel2013}, we observe mesoscale turbulence, characterized by vortical flow structures with a typical length scale $\Lambda_\Gamma \approx 2 \pi \sqrt{\Gamma_2/\Gamma_0}=2 \pi$. In contrast, at higher activity ($\A=1.1,\lambda=1.7$), the instantaneous  velocity field  (see \cref{fig:f1}(b)) clearly shows the emergence of an ordered state with non-zero  average velocity $\Vvec(t)=\langle \uvec \rangle$, where the angular brackets denote spatial averaging. However, the pseudocolor plot of the magnitude of the vorticity field in a plane perpendicular to $\Vvec$ reveals the presence of background turbulence. We refer to this new flow regime as a turbulent flock.

\begin{figure*}
\includegraphics[width=0.48\linewidth]{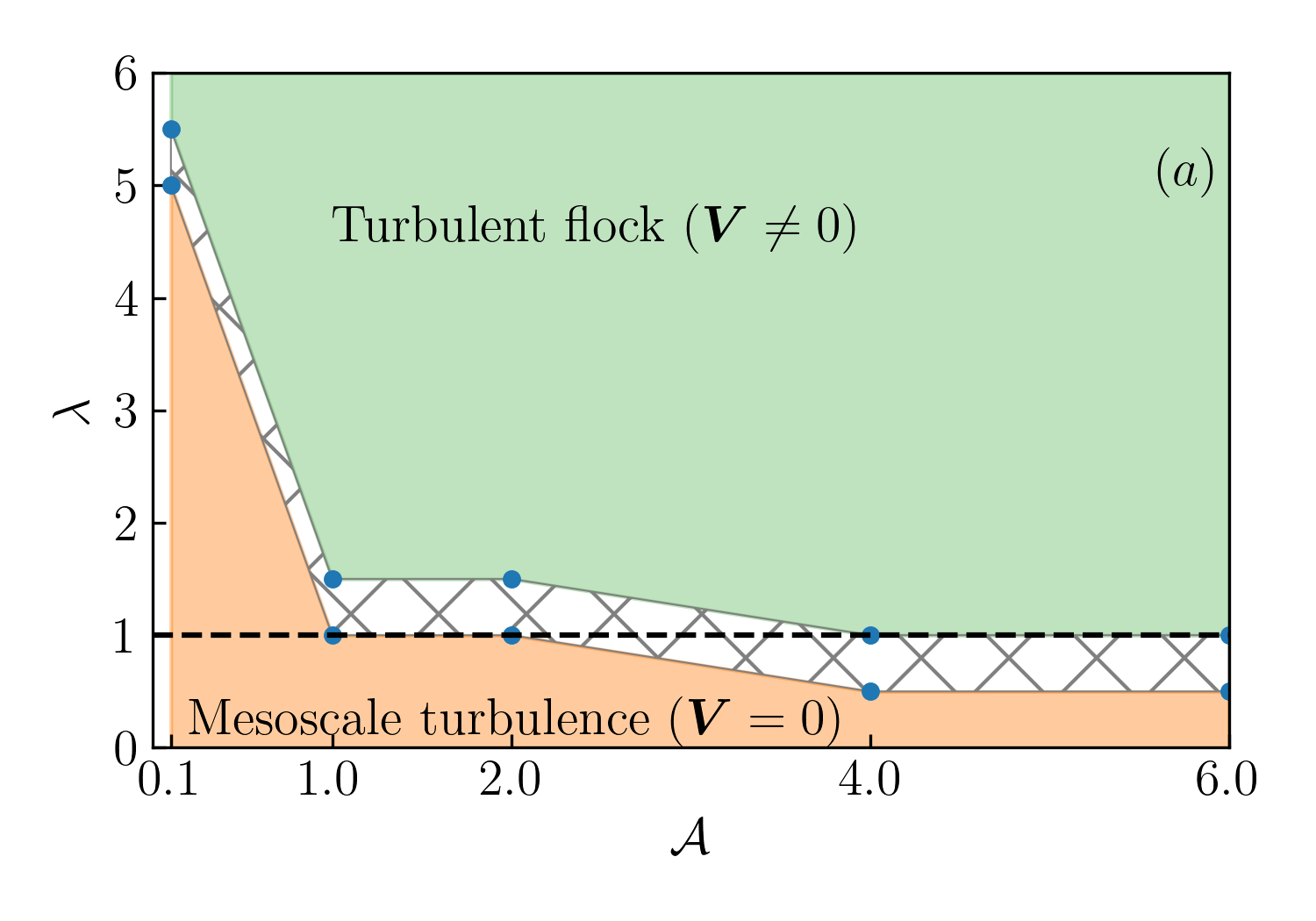}
\includegraphics[width=0.48\linewidth]{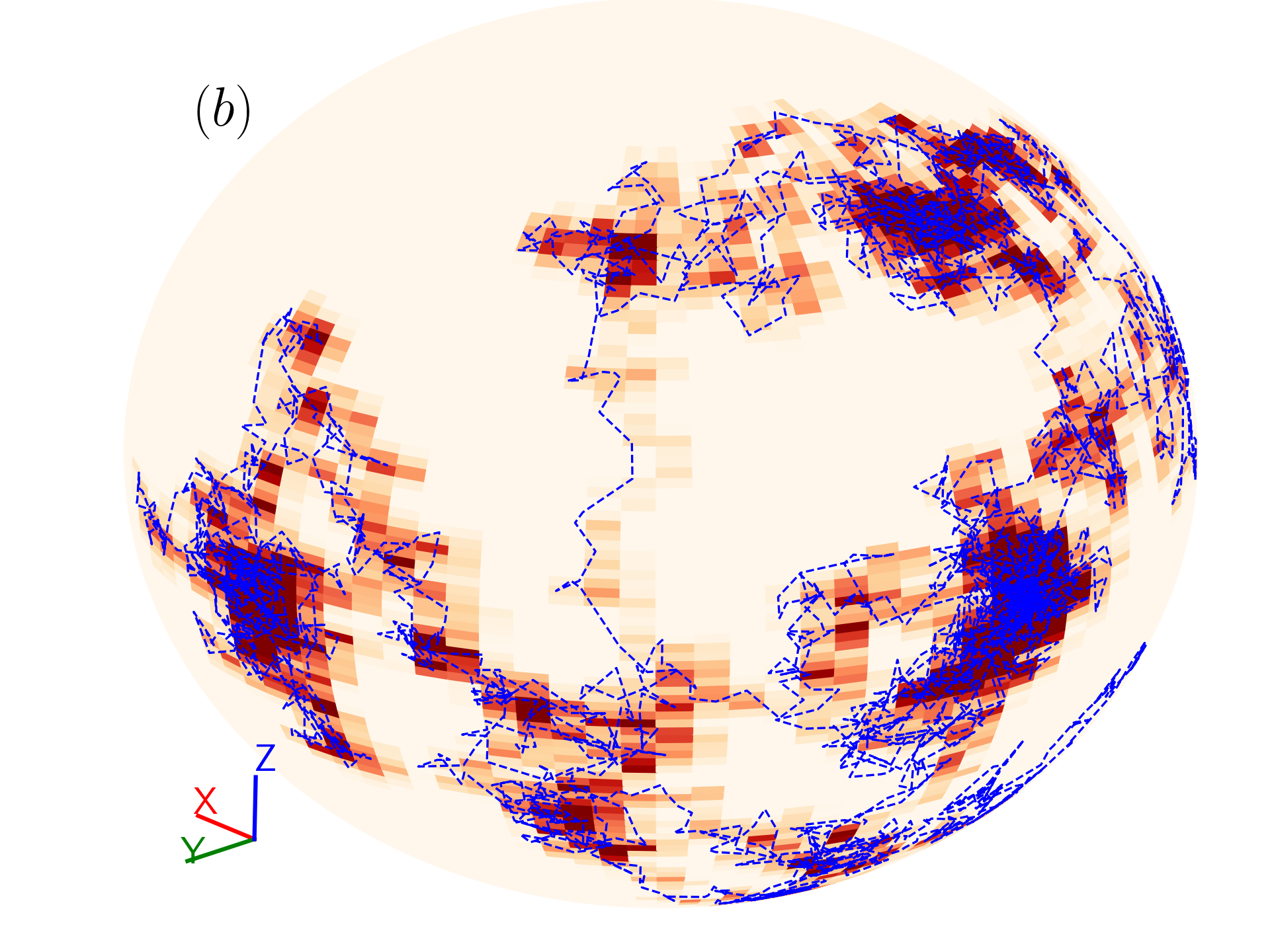}\\
\includegraphics[width=0.48\linewidth]{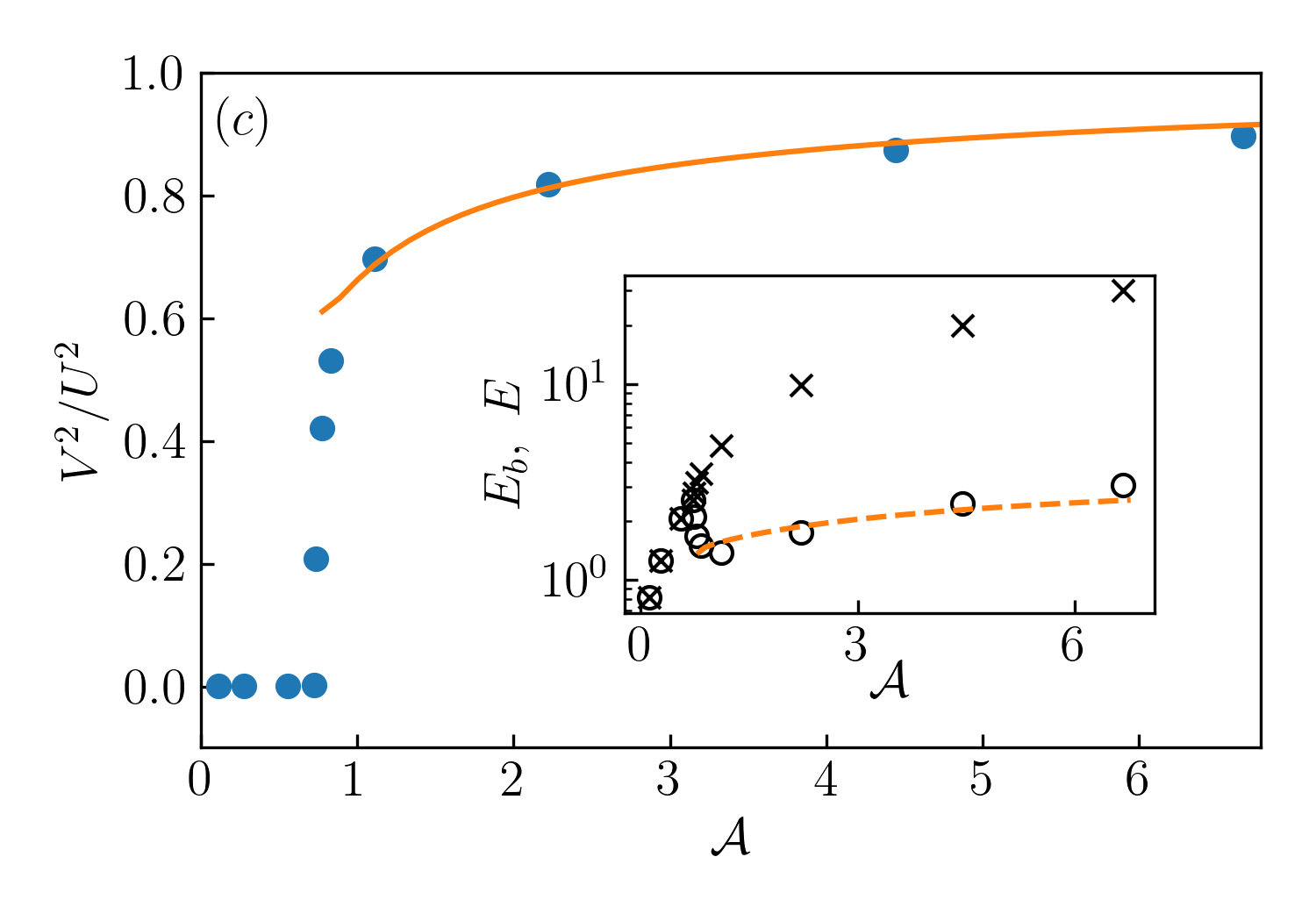}
\includegraphics[width=0.48\linewidth]{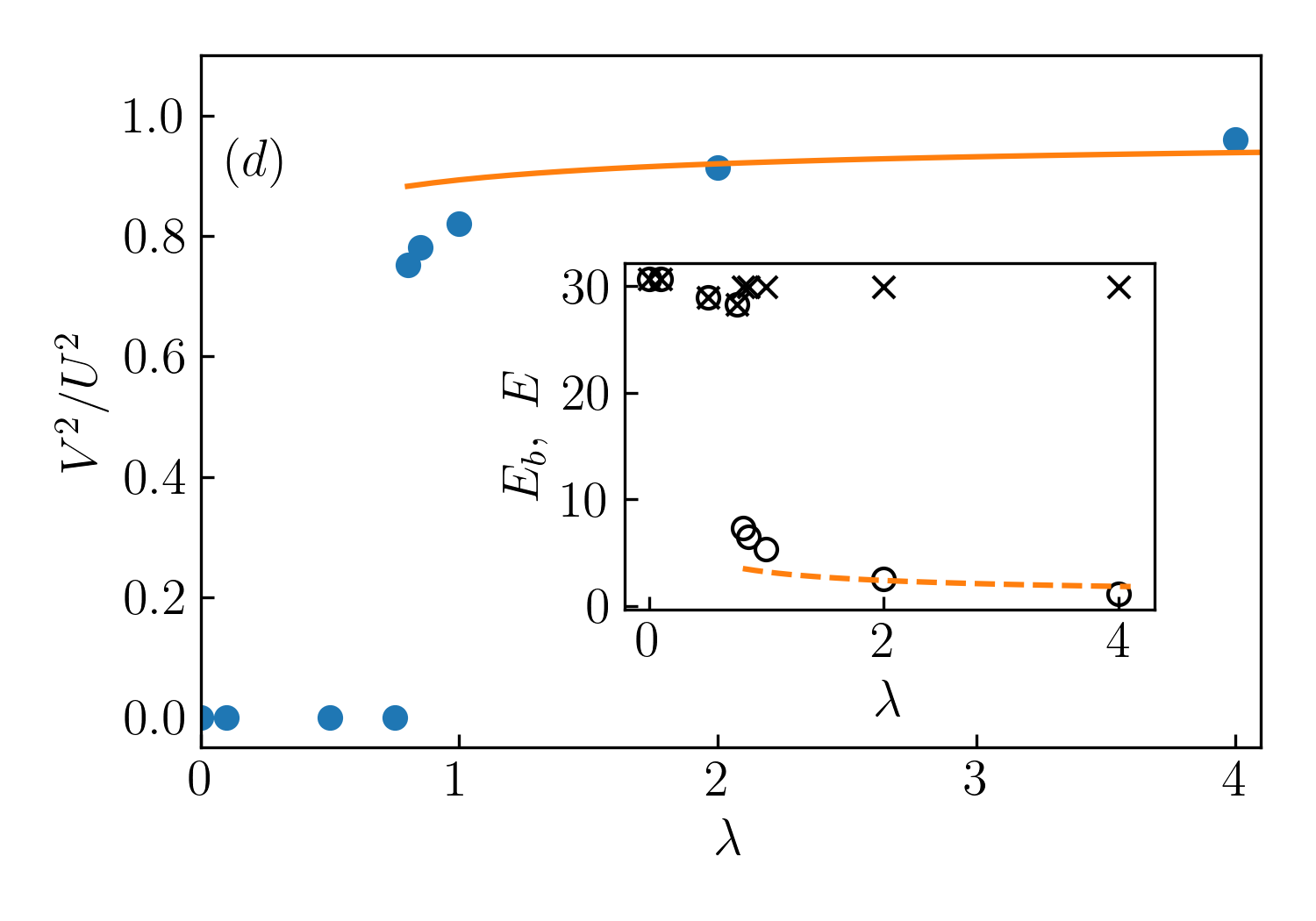}
\caption{\label{fig:meanu}(a) Numerically obtained phase-diagram showing regions of mesoscale turbulence and turbulent flock. The hash-region denotes the cross-over region, and  the  blue dots mark the simulations used to identify the phase boundary (runs {\tt R3-R4}). (b) A typical ${\evec}_{\Vvec}$ trajectory in the turbulent flock regime, and the density map showing the time spent by the trajectory about an orientation (dark red regions indicate longer time). (c) Plot of $V^2/U^2$ vs. $\A$ with fixed $\lambda=1.7$ (runs {\tt R1}). A transition from mesoscale to flocking  turbulence appears around $\A=0.73$. (Inset) Plot of $E_b$ vs. $\A$ (open circle) and $E$ vs. $\A$ (cross). (d) Plot of $V^2/U^2$ versus $\lambda$ for fixed $\A=6.6$ showing  transition from mesoscale to turbulent flock around $\lambda=0.86$ (runs {\tt R2}). (Inset) Plot of $E_b$ vs. $\lambda$ (open circle) and $E$ vs. $\lambda$ (cross). The orange lines in (b,c) show that the predictions for $(E_0,E_b)$ from closure theory \cref{eq:steady1} are in excellent agreement with the data.}
\end{figure*}

We use the magnitude of the average velocity ${V=\overline{({\Vvec}\cdot \Vvec)}^{1/2}}$, as the order parameter to characterize the transition from mesoscale turbulence ($V=0$) to turbulent flock ($V\neq0$). Here, the overline indicates a temporal average in the statistically steady state. The phase diagram in \cref{fig:meanu}(a) shows the crossover between these two regimes and is constructed from multiple DNS runs ({\tt R1}--{\tt R4} in \cref{table}). We have confirmed that these results do not depend on the system size ($L=24\pi$ or $L=96\pi$). For a fixed $\A$ ($\lambda$), flocking is observed for a large $|\lambda|\gg1$ ($\A\gg1$). In particular, for a fixed $\lambda=1.7$, a crossover from $V=0$ to $V\neq 0$ is observed around $\A=1$ (\cref{fig:meanu}(c)). Similarly, for a fixed $\A=6.67$, the transition between the two regimes occurs around $\lambda=0.86$ (\cref{fig:meanu}(d)). For large $\A$ or $\lambda$ in the turbulent flock regime, the magnitude of the velocity $V\approx U$.

We monitor the direction of the mean velocity vector ${\evec}_\Vvec \equiv{\Vvec}/|{\Vvec}|$ in the turbulent flock regime using a very long simulation of duration $t=5 \cdot 10^{5} \tau$ (run {\tt R3}). \Cref{fig:meanu} (b) displays the time evolution of ${\evec}_\Vvec(t)$ overlaid on a density map indicating how long the trajectory remains at each location. The orientation vector undergoes a random walk: it fluctuates around a given direction for some time and then escapes to adopt a new orientation.

A linear stability analysis of the TTSH equations  shows that the homogeneous solutions ($\uvec=0$ and $\uvec={\bm U}$) are unstable  \cite{wensink2012}. In the following, we motivate and show that the novel turbulent flock regime appears because of the nonlinear mechanisms that couple the mean flow and fluctuations.

Using \cref{eq:ttsh}, we obtain the following equation for the evolution of the total kinetic energy:
\begin{align}
\frac{dE}{dt} = 2 \alpha E + \Gamma_0 \langle |\nabla \uvec|^2 \rangle - {\beta} \langle |\uvec|^4 \rangle  - \Gamma_2 \langle |\nabla^2 \uvec|^2 \rangle, 
\label{eq:KE}
\end{align}
where $E= \langle\uvec^2 \rangle/2$. On the right-hand side, the first two terms inject energy, while the last two terms dissipate. Further decomposing the velocity into the mean velocity and a fluctuating part $\bvec=\uvec-{\bm V}$, we get $E=E_0 + E_b$  with $E_0=V^2/2$, and $E_b=\langle \bvec^2 \rangle/2$. 

In \cref{fig:meanu}(c,inset), we show that for a fixed $\lambda=1.7$ the total energy $E$ increases monotonically with $\A$. In the mesoscale turbulence regime ($\A<0.73$), $E_0=V\approx 0$ and $E_b\approx E$. Just above the cross-over, part of the energy is transferred to the mean flow, resulting in a decrease in $E_b$ and an increase in $E_0=V^2/2$. Finally, as expected, on further increasing $\A$, both the energy content of the mean flow $E_0 \propto V^2/2$ and $E_b$ increases.

Similarly, in \cref{fig:meanu}(d, inset) we study the variation in $E_b$ with $\lambda$ for a fixed $\A = 6.67$. Here, the total energy remains essentially constant at $E \approx U^2/2=30$ independently of $\lambda$. In the mesoscale turbulence regime ($\lambda < 0.86$), we find $E_b \approx E$. Around $\lambda = 0.86$ there is a sharp cross-over, beyond which, in the turbulent flock regime, the energy is predominantly stored in the mean flow, $E \approx E_0$, while $E_b$ decreases monotonically with increasing $\lambda$. Interestingly, and contrary to canonical turbulent flows \cite{Fri96}, strengthening the advective nonlinearity in \cref{eq:ttsh} actually facilitates the emergence of a flocking state (large-scale order).

We now derive a closure model in terms of the mean and the fluctuating energy to understand the transition. By spatial averaging \cref{eq:ttsh}, we obtain the equation for the evolution of the mean velocity:
\begin{align}
\frac{d\Vvec}{dt} = {\alpha} \Vvec - {\beta} \langle |\uvec|^2 \uvec  \rangle.
\label{eq:mvel}
\end{align}

Using \cref{eq:ttsh,eq:KE,eq:mvel}, we obtain the following equations for $E_o$ and $E_b$:
\begin{align}
\frac{dE_0}{dt}&= 2 \alpha E_0 - \beta \langle |\uvec|^2 \uvec  \rangle \cdot \Vvec,~{\rm and}\\
\frac{dE_b}{dt}&=  2 \alpha E_b - \beta \langle |{\bm u}|^2 {\bm u} \cdot {\bm b} \rangle + \varepsilon 
\label{eq:EE}
\end{align}
where $\varepsilon=\Gamma_0 \langle |\nabla \bvec|^2 \rangle - \Gamma_2\langle |\nabla^2 \bvec|^2 \rangle$ is the net injection (dissipation) due to small scale stirring. 

 For closure, we model the components of $\bvec$ as an uncorrelated random variable with vanishing odd-moments. Our DNS analysis reveals that in the flocking phase, fluctuations along the alignment directions are markedly reduced, specifically $\langle {\bvec \cdot {\bm e}_{\bm V}} \rangle \ll \langle \bvec \cdot {\bm e}_\perp \rangle$, and $\langle {\bm b}^4 \rangle \sim \langle {\bm b}^2 \rangle^2$, with ${\bm e}_\perp \cdot {\bm e}_{\bm V}=0$.   

Using the assumptions above, we get 
\begin{subequations}
\begin{align}
\langle |\uvec|^2 \uvec  \rangle \cdot \Vvec &\approx 4 E_0^2 \left(1 + \frac{E_b}{E_0}\right), \label{eq:approx1} \\
%\label{eq:approx1}
%\end{align}
%\begin{align}
\langle |\uvec|^2 \uvec \cdot \bvec \rangle &\approx 4 E_0^2 \left(\frac{E_b}{E_0} + C \frac{E_b^2}{E_0^2} \right).
\label{eq:approx2}
\end{align}
\label{eq:approx}
\end{subequations}

\begin{figure}[!h]
\includegraphics[width=\linewidth]{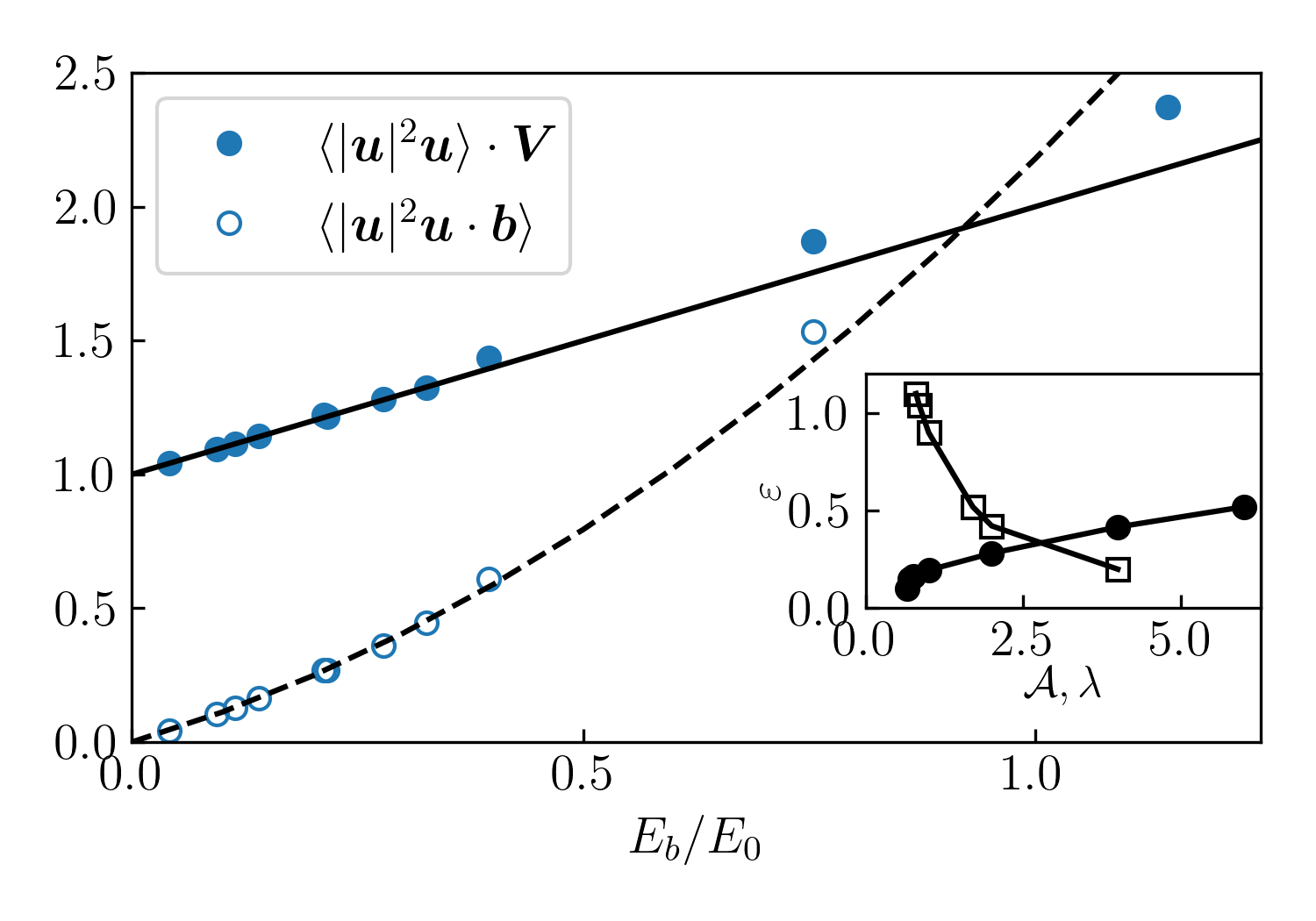}
\caption{\label{fig:bal} Plots of $\langle {\bm u}^2 {\bm u} \rangle \cdot {\bm V}/4 E_0^2$ (filled circle) and $\langle {\bm u}^2 {\bm u}  \cdot {\bm b}\rangle/4 E_0^2$ (open circle) versus $E_b/E_0$. The curves result from a least squares fitting (utilizing approximations \cref{eq:approx1,eq:approx2}) applied to data within the flocking regime, defined by $V^2/U^2 \geq 0.8$. The solid black line shows the plot $1+E_b/E_0$ (rhs of \cref{eq:approx1}), whereas the dashed black line corresponds to the fit $E_b/E_0 (1+ C E_b/E_0)$ (rhs of \cref{eq:approx2}) with $C=1.17\approx 6/5$.
(Inset) Plots showing  $\varepsilon$ vs. $\A$  (circle, fixed $\lambda=1.7$), and $\varepsilon$ vs. $\lambda$ (circle, fixed $\A=6.6$).}
\end{figure}

The plots in \cref{fig:bal} show the excellent agreement between  $\langle |{\bm u}|^2 {\bm u} \rangle\cdot {\bm V}$ and $\langle |{\bm u}|^2 {\bm u} \rangle\cdot {\bm b}$, and their approximations (rhs of \cref{eq:approx1} and \cref{eq:approx2}). The parameter $C\approx 6/5$ is obtained using the least squares method. As expected, departures are observed in the cross-over regime $0.1< V^2/U^2 <0.8$.

Substituting \cref{eq:approx} into \cref{eq:EE}, we obtain
\begin{equation}
\begin{aligned}
\frac{d E_0}{dt}&= 2 ({\alpha} - 2 \beta  E_b) E_0 - 4 \beta E_0^2,~{\rm and} \\
\frac{dE_b}{dt} &= 2 \alpha E_b -4 \beta \left(  E_b E_0  +  \frac{6}{5} E_b^2\right) + \varepsilon.
\label{eq:closure}
\end{aligned}
\end{equation}

The value of $\varepsilon$ depends on the activity parameters $\A$ and $\lambda$. For example, in \cref{fig:bal}(inset) we show that   $\varepsilon$ monotonically decreases (increases) with increasing $\lambda$ ($\A$) for fixed $\A=6.6$ ($\lambda=1.7$). We treat $\varepsilon$ as an independent control parameter in our closure model.

\begin{figure}[!h]
\includegraphics[width=\linewidth]{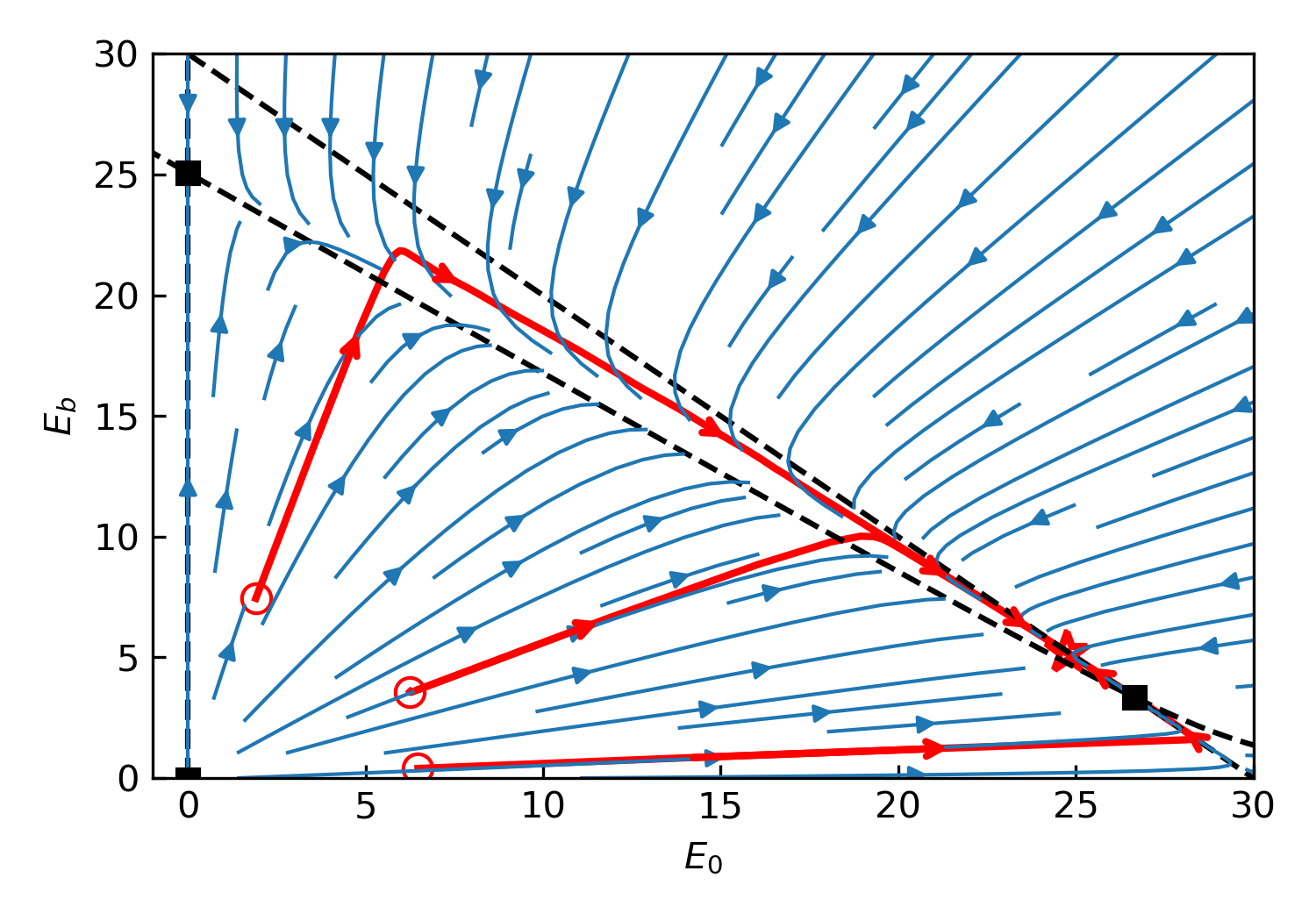}
\caption{\label{fig:phase} Phase portrait showing the dynamics in the positive $(E_0,E_b)$ quadrant obtained using \cref{eq:closure} with $\A=6.6$, $\B=0.09$, and $\lambda=1.7$. The fixed points are marked with black squares and the nullclines are shown with black dashed lines. The fixed point $(0,0)$ is a unstable-node, the non-zero $(E_0,E_b)$ \cref{eq:steady1} is a stable node, and $(0,E_b)$ \cref{eq:steady2} is a saddle. The red trajectories, obtained from DNS for different initial conditions (circles), all terminate near a star-marked point that is located close to the stable node \eqref{eq:steady1}.}
\end{figure}

The steady-state solutions of \cref{eq:closure} in the physically relevant regime of $E_0>0,E_b>0$ are as follows:
\begin{subequations}
\begin{align}
(E_0,E_b)&=\left(\frac{U^2}{2}-\sqrt{\frac{5 \epsilon}{4 \beta}}, \sqrt{\frac{5 \epsilon}{4 \beta}} \right),~\rm{and} \label{eq:steady1}\\
%\end{align}
%\begin{align}
(E_0,E_b)&=\left(0,\frac{5U^2}{25} \left[1 +\sqrt{1 + \frac{24\epsilon}{5 \beta U^4}} \right]\right).
\label{eq:steady2}
\end{align}
\end{subequations}
The solution \cref{eq:steady1} represents a state in which  large-scale order and turbulence coexist, resulting in a dynamic flock. In contrast, \cref{eq:steady2} pertains to the extensively studied mesoscale turbulence regime \cite{wensink2012,dunkel2013}. A linear stability analysis of \cref{eq:closure} reveals that when $\varepsilon< \varepsilon_c$, where $\varepsilon_c\equiv \beta U^4/5$, the fixed point \cref{eq:steady2} is a saddle, while \cref{eq:steady1} is a stable node. The phase portrait shown in \cref{fig:phase} confirms this behavior. By superimposing our DNS-derived $(E_0,E_b)$ trajectories on the phase portrait for various initial conditions, we observe strong agreement. In particular, steady-state values $(E_0,E_b)$ are proximate to the fixed point \cref{eq:steady1}. In \cref{fig:meanu}(c,d), we quantitatively verify that the steady-state $(E_0,E_b)$ obtained from our closure model is in close agreement with the DNS values for different values of $\A$ and $\lambda$ in the flocking regime.

Furthermore, as $\varepsilon$ increases, the stable fixed point approaches the saddle point. At $\varepsilon = \varepsilon_c$ the fixed points merge. A further increase in $\varepsilon$ leads to their crossing, with a subsequent stability exchange through a transcritical bifurcation. For $\varepsilon > \varepsilon_c$, the fixed point described by \cref{eq:steady2} that corresponds to the mesoscale turbulence regime stabilizes. 

Consider \cref{fig:meanu}(d), where we fix $\A=6.6$, and vary $\lambda$ (or equivalently $\varepsilon$, see \cref{fig:bal}(inset)). At the  cross-over ($\lambda=0.75$) from the mesoscale turbulence to the turbulent flock   $\varepsilon\approx 1.1$ which is much smaller than $\varepsilon_c=72$ predicted by the closure. This discrepancy is not surprising, as the approximations used to derive the closure model \cref{eq:closure} are only valid deep in the flocking regime (large values of $\A$ and $\lambda$).

\begin{figure}[!h]
\includegraphics[width=\linewidth]{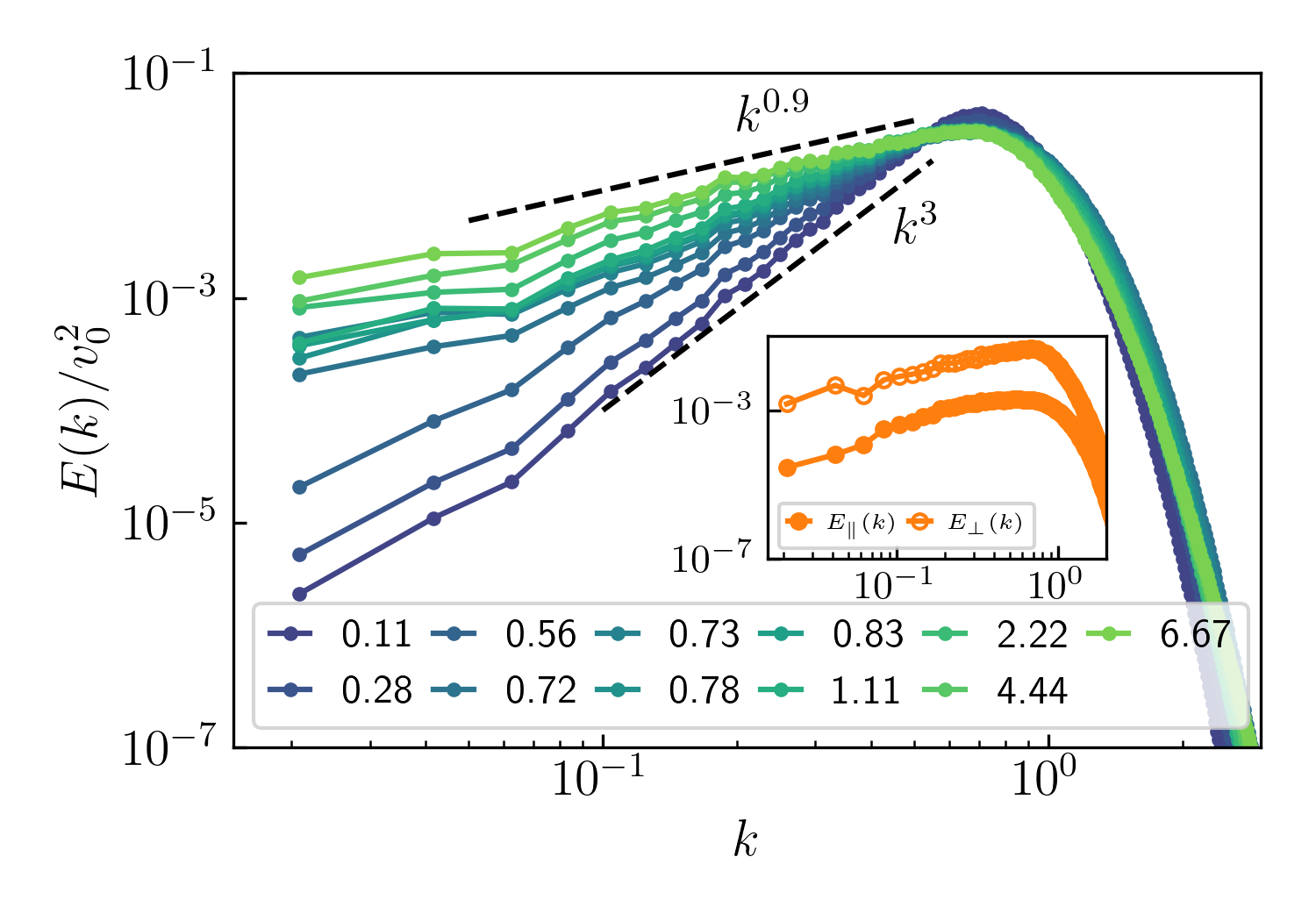}
\caption{\label{fig:spec} Energy spectrum for different values of $\A$ (see legend) while keeping fixed $\B=0.09$ and $\lambda=1.7$. At low $k$, we observe $E(k)\sim k^\delta$, where the exponent $\delta$ decreases from $-3.2$ to 1 as $\A$ increases \cite{bratanov2015,sanjay2020}. The inset shows components of the spectra in the flocking state ($\A=6.6$): $E_\parallel$ corresponds to the components of $\bvec$ aligned with the order direction ${\bm V}/{|\bm V|}$, and $E_\perp$ represents a direction orthogonal to it. As pointed out in the approximation for the closure model, we observe the transverse fluctuations dominate, $E_\|\sim 25 E_\perp$.}
\end{figure}

As is typical for fluid turbulence, we characterize the flow fluctuations in the two regimes using the energy spectrum $E(k)=\int |{\bm b}({\bm q})|^2 \delta(k-{|\bm q|})d{\bm q}$ of the flow fluctuations with varying activity parameter ${\A}$, while $\B=0.09$ and $\lambda=1.7$ are fixed (\cref{fig:spec}). For small $k$, we recover the characteristic power-law scaling ($E(k)\sim k^\delta$) of active turbulence \cite{bratanov2015,sanjay2020,mukherjee2023} in both the mesoscale turbulence and turbulent flock regimes; the exponent $\delta$ decreases from $3$ to $0.9$ as $\A$ increases. However, in a turbulent flock, as shown in \cref{fig:spec}(inset), fluctuations along the mean flow direction are strongly suppressed compared to transverse fluctuations, while they remain isotropic in the mesoscale turbulence regime ($\A\leq 0.65$) \cite{dunkel2013}.

{\textit{Conclusions}} -- We report a novel turbulent flock regime for large activity ($\lambda\gg 1$ or $\A\gg1$) in the three-dimensional TTSH equations, a minimal model for turbulence in dense bacterial suspension in the Stokesian regime ($\Rey \to 0$). We present a closure model in terms of the energy contained in the mean flow and fluctuations and validate it against DNS. Not only does the model correctly predict a stable turbulent flock regime, it also qualitatively captures the transition from turbulent flock to mesoscale turbulence on reducing activity. Our results are also consistent with recent experiments in bacterial suspensions, which show that the size of the coherent flow structure increases with a reduction of confinement \cite{wei2024}, suggesting the possibility of global order in unbounded domains.

We emphasize that our result on turbulent flock is in the Stokesian regime ($\Rey \to 0$), in contrast to recent studies that observe flocking in the inertial regime ($\Rey \to O(1)$) \cite{chatterjee2021,rana2022,jain2024,jain2025}.

Finally, we note that, for large activity, the turbulent flock regime in 3D differs markedly from its 2D counterpart which shows large-scale vortical structures interspersed with mesoscale eddies, and an inverse energy cascade \cite{mukherjee2023,kiran2025,kashyap2025}. These results underscore, in close analogy to conventional fluid turbulence, the central role of advective nonlinearities. We anticipate that our work will stimulate further experimental and theoretical studies of how nonlinear interactions and confinement jointly shape the statistical properties of turbulence in swimmer suspensions.

We acknowledge support from the Department of Atomic Energy (DAE), India under Project Identification No. RTI 4007,  DST (India) Project No. MTR/2022/000867 and the National Supercomputing Mission (NSM) for providing
computing resources of ‘PARAM Siddhi-AI’ at C-DAC, Pune which is implemented by C-DAC and supported by the Ministry of Electronics and
Information Technology (MeitY) and the Department of Science and Technology
(DST), Government of India. All simulations are performed using the Turing GPU cluster at TIFR-Hyderabad and on the PARAM Siddhi cluster at CDAC-Pune. We  acknowledge stimulating discussions, support, and hospitality during the program `Festival de Théorie-2025' in Aix-en-Provence.

\appendix
\section{End Matter}
\subsection{Dimensionless equations}
 We define the typical length scale $\ell\equiv \sqrt{\Gt/\Gz}$, and the velocity scale $\vo \equiv \sqrt{\Gz^3/\Gt}$ \cite{bratanov2015}. Rescaling space ${\xvec} \to {\xvec/\ell}$, time $t\to t/\tau$, and pressure $P \to P/\vo^2$,  we get 
\begin{align*}
\partial_t \uvec + \lambda  \uvec \cdot \nabla \uvec = -\nabla P - \nabla^2 \uvec -  \nabla^4 \uvec + \left(\A - \B |\uvec|^2 \right) \uvec,
\end{align*}
where $\A\equiv \alpha \Gt/\Gz^2$, and $\B\equiv \beta \Gz$.   

\subsection{Simulation Parameters}
The parameters used in our simulations are summarized in Table~\ref{table}.
\begin{table}[H]
    \centering
    \setlength{\tabcolsep}{3pt}
    \begin{tabular}{lcccc}
     \hline \hline
     ${\tt RUN}$ & $L$ & $N$ & $\lambda$ & $\alpha$   \\ 
       \hline
       \multirow{2}{*}{\tt R1} & \multirow{2}{*}{$96 \pi$} & \multirow{2}{*}{$1024$}  & \multirow{2}{*}{$1.7$} &   {$0.1,0.25,0.5,0.65,0.66,0.7,$}   \\ 
                                &    &    &  & {$0.75,1.0,2.0,4.0,6.0$}  \\       
       \multirow{2}{*}{\tt R2} & \multirow{2}{*}{$96 \pi$} & \multirow{2}{*}{$1024$}  & {$0.1,0.5,0.75,0.8,$}  &  \multirow{2}{*}{$6.0$}   \\ 
                                &    &  & {$0.85,1.0,2.0,4.0$} &    \\
       {\tt R3} &{$24 \pi$} & {$256$}  & {$4.0,5.0,5.5,6.0$}  & {$0.1$}   \\ 
       {\tt R4} &{$24 \pi$} & {$256$}  & {$0.5,1.0,1.5,2.0$}  & {$1.0,2.0,4.0,6.0$}   \\
    \hline \hline
    \end{tabular}
    \caption{\label{table} The parameter values used in our DNS. We fix $\beta=0.1$, $\Gz=0.9$, and $\Gt=0.9$ \cite{dunkel2013}.}
\end{table}

\end{document}